# Influence of water intercalation on the electronic structure of the hydrated $Na_{0.3}CoO_2 \cdot yH_2O$ using a local spin density approximation

R. J. Xiao, H. X. Yang and J. Q. Li*

Beijing National Laboratory for Condensed Matter Physics, Institute of Physics, Chinese Academy of Sciences, Beijing, 100080, P.R. China

The electronic band structures of the parent compound $Na_{0.3}CoO_2$ and the hydrated superconductor $Na_{0.3}CoO_2 \cdot 1.3H_2O$ were evaluated using the local spin density approximation (LSDA). Systematic analysis on the bands near the Fermi surface (FS) revealed that the hydration-induced modifications appear, notably in the valence charge density and the orbital hybridization. The most striking changes are the decrease of density in Cobalt's $a_{1g}$ state and the rehybridization between O $2p$ and Co $a_{1g}$ in the hydrated superconducting phase. These hydration-induced modifications could yield visible differences in interband excitations and dielectric functions. Experimental results of electron energy-loss spectroscopy (EELS) are fundamentally in agreement with the theoretical predictions for the hydrated $Na_{0.3}CoO_2 \cdot yH_2O$.

The recent discovery of superconductivity in $Na_{0.3}CoO_2 \cdot 1.3H_2O$ [1] has stimulated extensive interest in $Na_xCoO_2$ systems, known previously as valuable materials with high thermopowers [2]. In order to understand the origin of superconductivity and other related physical properties in $Na_{0.3}CoO_2 \cdot yH_2O$ (y=0, 0.6, 1.3) and $Na_xCoO_2$ (0.3<x<0.8) materials, numerous theoretical studies have been performed to investigate the effects of $Na/H_2O$-intercalation on the electronic structure. The band calculations revealed that the octahedral crystal-field causes a splitting of ~2.5eV between the $e_g$ and $t_{2g}$ states in $Na_xCoO_2$, and that the $t_{2g}$ manifold splits into $a_{1g}$ and $e'_g$ bands; the Fermi level ($E_F$) lies within the spin-down $t_{2g}$ states and forms an FS consisting of a large cylinder and six small pockets [3-5]. In addition, Coulomb correlation was also demonstrated to be an important factor for determining the electronic structure and the FS topology in $Na_xCoO_2$ (0.3<x<0.75) [4, 6, 7]. The water intercalation can produce a series of hydrated phases [8]. The three notable ones are known as the metallic $Na_{0.3}CoO_2$ phase, the intermediate $Na_{0.3}CoO_2 \cdot 0.6H_2O$ phase and the superconducting $Na_{0.3}CoO_2 \cdot 1.3H_2O$ phase [9]. According to the results of the first principles calculations for $Na_xCoO_2 \cdot 1.3H_2O$ [10], the $H_2O$ intercalation in general makes the electronic structure more two-dimensional and does not lead to notable differences in the band structure near the FS. On the other hand, a variety of experimental investigations have suggested that the transport properties and magnetization change notably along with water intercalation [11]. Moreover, our recent EELS measurements have revealed a systematic change in the spectra along with water intercalation [12], the remarkable changes appears below 10eV originating chiefly from collective plasmons and inter/intra-band transitions, these transitions are essentially governed by the electronic band structure near the $E_F$. In this paper, we have systematically analyzed the band characteristics, the valence charge density and orbital hybridization in $Na_{0.3}CoO_2 \cdot yH_2O$ materials. Remarkable modifications arising from hydration are noted at the charge density of the Co $a_{1g}$ state and orbital hybridization between O $2p$ and Co $a_{1g}$ states. Theoretical results on energy-loss functions are in good agreement with experimental measurements.

In our theoretical calculations, we used the full potential linearized augmented-plane-wave+local orbital (APW+LO) model as implemented in WIEN2k code [13]. It is known that the accurate structure of the $H_2O$ clusters among the $CoO_2$ sheets is a very complex issue in the study of $Na_{0.3}CoO_2 \cdot 1.3H_2O$ superconductor, the coexistence of the $H_2O$ and $H_3O^+$ clusters is suggested based on the experimental results [14]. On the other hand, the band structural calculations indicated that the $H_2O$ intercalation does not lead to notable differences on the bands near the FS [10]. Hence, in present study, we will use a simplified model without specific considerations of the $H_2O$ structure as discussed in following context. Lattice expansion arising from water intercalations are cited from the experimental data for $Na_{0.3}CoO_2 \cdot yH_2O$ materials [9]. The excess electrons were balanced by a uniform positive background [15]. The height of O is respectively relaxed for each model with different water content. The dielectric functions were calculated using the OPTIC program of the WIEN2k distribution. In our EELS experiments, we performed the measurements on numerous single-crystalline $Na_{0.3}CoO_2$ samples and polycrystalline $Na_{0.3}CoO_2 \cdot yH_2O$ (y = 0, 0.6 and 1.3) samples. They were all well characterized in the previous structural and physical measurements [16].

Considering that the spin-up bands do not cross the $E_F$ [3], we first performed our analysis on the valence charge density of the spin-down states. $Na_xCoO_2$ has a $D_{6h}$ point group, it is convenient to choose the threefold symmetry axis as the $z$-direction in calculations. In this coordinate system, the $a_{1g}$ state extends along the $c$ axis, and the $e'_g$ states spread over the $a$-$b$ plane. Figure 1 shows the charge density difference between $Na_{0.3}CoO_2$ and $Na_{0.3}CoO_2 \cdot 1.3H_2O$, the data shown was directly obtained from the local electron density of $Na_{0.3}CoO_2·1.3H_2O$, subtracting that of $Na_{0.3}CoO_2$. The striking feature recognizable in this figure is the lower density of Co $a_{1g}$ state (note the green region in Fig.1c) in the superconducting phase. Moreover, changes of electron density around oxygen atoms are also analyzed – i.e., a clear increase between the O and Co layers and a decrease between the O and Na layers (Fig. 1a and b). It is known that the water intercalation directly enlarges the distance between $CoO_2$ layers and decreases interplanar interaction in the hydrated phase. As a result, the O $2p$ electrons between the O and Na layers transfer partially to the space between O and Co atoms – this alteration leads to the orbital rehybridization between O $2p$ and Co $a_{1g}$ states. In fact, hole density increases in the $a_{1g}$ state, and electron density is added to the oxygen states. Meanwhile we also performed a calculation to reveal the hydration-induced changes in the $e'_g$ states, with results suggesting that these states show much less change in comparison with the $a_{1g}$ state. The occurrence of orbital rehybridization between O $2p$ and Co $a_{1g}$ states in the hydrated phase may be important as similarly discussed in the studies of high-$T_c$ superconductors, in which the hybridization of the Cu $3d$ with O $2p$ orbitals were particularly considered for interpreting certain remarkable phenomena – e.g. superconductivity, electronic phase separation and charge stripes [17]. In recent literature, the mechanism for the superconductivity in $Na_{0.3}CoO_2·1.3H_2O$ has been studied based on several models, including the extended $s$-wave pairing that originates from the large $a_{1g}$ cylinder FS [18] and the triplet $f$-wave pairing that originates from the six $e'_g$ pockets FS [19]. According to our analysis of the hydration-induced alternation of valence charge density, the $a_{1g}$ state is possibly an important factor in understanding the superconductivity in $Na_xCoO_2·yH_2O$.

    Based on the calculated band structures for $Na_xCoO_2$ and $Na_xCoO_2·1.3H_2O$ [10], we further calculated the dielectric functions, the interband excitations and then the energy-loss

functions [20], which can be directly compared with the experimental EELS data. In our calculation the number of $k$-points was increased to 20,000 (~960−1700 irreducible depending on different structure sizes) for accuracy. The imaginary part $\varepsilon_2$ of the dielectric function was obtained by [20]:

$$\varepsilon_{2ii} = \frac{4\pi^2 e^2}{m^2(\omega - \Delta_c/\hbar)^2 V} \sum_{v,c,k} |\langle ck|p_i|vk \rangle|^2 \times \delta(E_{ck} - E_{vk} - \hbar\omega), \qquad (1)$$

where $E_{ck}$ and $E_{vk}$ are the quasiparticle energies approximated to the eigenvalues, $|ck\rangle$ and $|vk\rangle$ are the Bloch functions of the conduction and valence bands, which reveal certain changes along the water intercalation; $V$ is the volume of the unit cell, and $p_i$ is the momentum operator with $i=x$, $y$ or $z$ direction corresponding to the crystal Cartesian axes. Then, the Kramers-Krönig (KK) analysis is performed to obtain the real part $\varepsilon_1$ and the loss function Im(-1/$\varepsilon$).

In order to clearly illustrate the anisotropic properties in the $Na_{0.3}CoO_2$ materials, the imaginary part $\varepsilon_2$ and real part $\varepsilon_1$ of the dielectric function (the loss function as well) are respectively calculated for the directions parallel and perpendicular to $c$-axis as shown in Fig. 2. The main peaks in Im(-1/$\varepsilon$) (No.1 to No.6) are attributed to the interband transitions as indicated in Fig. 2b. The notable anisotropic features in the present system are recognizable as an apparent difference between $\varepsilon_{2xx}$ and $\varepsilon_{2zz}$; three transitions between 5 to 8eV can be observed in both $\varepsilon_{2xx}$ and $\varepsilon_{2zz}$, and the other three peaks (~2.5eV, 3.3eV and 4.8eV) appear only in the $x$ orientation. The inset of Fig. 2a shows the typical bands included in the interband transitions, arrows are positioned at the $k$-points contributing principally to the strong excitations. The strongest peak (No. 6) in the loss function is identified as a transition occurring near the K points from the bottom of O 2$p$ bands to the Co $e_g$ bands. The excitations in $\varepsilon_{2xx}$ arise chiefly from $p_x+p_y$ (O) to $s$ (O) and $d_{x^2+y^2}+d_{xy}$ (Co) to $p_x+p_y$ (Co), and the excitations in $\varepsilon_{2zz}$ come mostly from $p_x+p_y$ (O) to $d_{xz}+d_{yz}$ (O) and $p_z$ (O) to $s$ (O).

Similar theoretical analysis has also been performed for several water-intercalated phases with different $c$-axis parameters. The calculations on the $Na_{0.3}CoO_2 \cdot 1.3H_2O$ superconductor yield the similar peaks in the imaginary part $\varepsilon_2$, but its magnitude (intensity of the peaks) is much lower than that in the parent phase. Figure 3a shows the calculated average loss functions for the superconducting phase y=1.3 in the energy range of 2 to 12eV. Data for the parent phase y=0 are also shown for comparison. It is noted that spectral peaks in the y=1.3 phase shift visibly towards the lower energy direction with respect to that in the y=0 phase. In order to make a direct comparison with our experimental EELS data, the resolution of the theoretical data was lowered to 1eV as shown in Fig. 3b. Figure 3c shows the experimental spectra for y=0, 0.6 and 1.3 phases, as discussed to some extent in our previous paper [12]. The most striking feature, in good agreement with theoretical prediction, is the noticeable shift of main peaks along with water intercalation. For instance, the strong peak at around 7.9eV in $Na_{0.3}CoO_2$ shifts to lower energy of 7.1eV in $Na_{0.3}CoO_2 \cdot 0.6H_2O$, and then to 6.2eV in the $Na_{0.3}CoO_2 \cdot 1.3H_2O$ superconductor. Careful examinations suggest that all experimental spectra also contain many fine features in addition to the strong excitations. Especially, the spectrum for the superconducting phase $Na_{0.3}CoO_2 \cdot 1.3H_2O$ reveals the presence of several clear small peaks/shoulders below 10eV (as typically indicated by arrows). In additional to our previous interpretations [12], the clear peaks at 3.3eV are identified as a transition from Co $t_{2g}$ to Co $e_g$ based on our calculation (labeled as peak 2 in Fig. 3a), the peak at 9eV is attributed to a $H_2O$ molecule orbital transition [21]. It should be pointed out that the electronic structures of $H_2O$ molecules is not included in our calculations, therefore the 9eV peak is not visible in the theoretical spectrum for $Na_{0.3}CoO_2 \cdot 1.3H_2O$.

$Na_{0.3}CoO_2 \cdot yH_2O$, dependent on the material processing, has a series of hydrated phases

with distinctive $H_2O$ ordering among the $CoO_2$ sheets – i.e. y=0, 0.3, 0.6, 0.9, 1.3 and 1.8 [9]. Our systematic studies of the dielectric functions and the energy-loss functions revealed a notable change in association with hydration. As a result, Fig. 4 shows a theoretical relationship between the energy of the strong excitation (broadening=1eV) and the water content in $Na_{0.3}CoO_2 \cdot yH_2O$. In our experiment, we successfully prepared three typical samples with y=0 (single crystal), y=0.6 and 1.3, and carried out systematic EELS measurements as illustrated in Fig. 4. It is recognizable that the theoretical and experimental results tend to coincide, though the experimental data are somewhat lower than theoretical ones. In order to reach a better agreement, we must consider more detailed structural factors in our theoretical analysis, especially the atomic ordering of sodium and $H_2O$ sheets among $CoO_2$ sheets.

It is also noted that several theoretical studies considering Coulomb correlations in present system could yield a better explanation of some physical properties such as the FS topology [7] and disproportionate charge [4,6]. We therefore have further performed an analysis of the energy-loss functions using the DFT+U approach. It is demonstrated that the results of U<4eV give rise to similar spectra that are qualitatively comparable with the experimental ones. On the other hand, the theoretical spectra for U>4 eV show very complex peaks in both the dielectric function and the loss function in sharp contrast with the experimental data for either the parent or hydrated phase. It seems that the Coulomb correlations (U) in $Na_{0.3}CoO_2 \cdot yH_2O$ (0 ≤ y ≤ 1.3) systems are smaller than 4eV. We have tried to improve the agreements between the experimental and theoretical data as shown in Fig. 4 by adjusting the Coulomb correlation (U), however, no visible improvement have been achieved in our DFT+U calculations. Similar facts were also noted in the study of optical properties [22], it is found that the DFT+U method cannot reproduce the experimental optical spectra of $Na_xCoO_2$ very well. The specific features of the intercalated layers, such as Na, $H_2O$, and $H_3O^+$ [14], could have certain influences on the low energy excitations in the present system. Further study is still in progress.

In summary, the electronic structures of the $Na_{0.3}CoO_2 \cdot yH_2O$ materials have been evaluated using the LSDA method. An extensive analysis of the bands near the FS revealed the presence of the remarkable hydration-induced modifications in the band characters, the valence charge density and orbital hybridization. The decrease of valence charge density in Cobalt's $a_{1g}$ state and the rehybridization between O $2p$ and Co $a_{1g}$ are possibly important factors in understanding the remarkable properties in the hydrated superconducting phase. The changes of electronic band structure in combination with the lattice expansion could result in a remarkable transformation of dielectric functions along with water intercalation. Experimental EELS results in the low energy range are fundamentally in agreement with the theoretical predictions for several typical samples of $Na_{0.3}CoO_2 \cdot yH_2O$ with x=0, 0.6 and 1.3.

## Acknowledgments

We would like to thank Prof. N.L. Wang for providing single-crystal samples of $Na_xCoO_2$ and Dr. L.F. Xu for very valuable discussions. The work reported here is supported by the National Natural Science Foundation of China.

Figure captions

Fig. 1 (color line) The spin–down charge density differences between $Na_{0.3}CoO_2\cdot 1.3H_2O$ and $Na_{0.3}CoO_2$. The shown planes cut through (a) $O_1$, (b) $O_2$ and $O_3$, and (c) Co atoms, respectively, illustrating the alternations of electron density along the $z$ direction. The units in the scale label are electrons/Å$^3$. (d) A schematic structural model for the $CoO_6$ octahedron.

Fig. 2 (a) Theoretical energy-loss function for y=0 perpendicular ($Im(-1/\varepsilon_{xx})$) and parallel ($Im(-1/\varepsilon_{zz})$) to the $c$ -axis. Inset: the interband transitions contribute respectively to the six main peaks in the energy-loss function. (b) Calculated real ($\varepsilon_{1xx}$, $\varepsilon_{1zz}$) and imaginary parts ($\varepsilon_{2xx}$, $\varepsilon_{2zz}$) for y=0.

Fig. 3 (a) Calculated average loss function for y=0 and 1.3 with broadening=0.1eV; (b) Calculated energy-loss function for y=0 and 1.3 with broadening=1.0eV; (c) Experimental EELS data for y=0, 0.6 and 1.3. Two small peaks in the y=1.3 spectrum are indicated by arrows.

Fig. 4 Relationship between the energy of the strong excitation and water content in $Na_{0.3}CoO_2\cdot yH_2O$: theoretical data for y=0, 0.3, 0.6, 0.9 and 1.3 (broadening=1.0eV), and experimental results for y=0, 0.6 and 1.3.

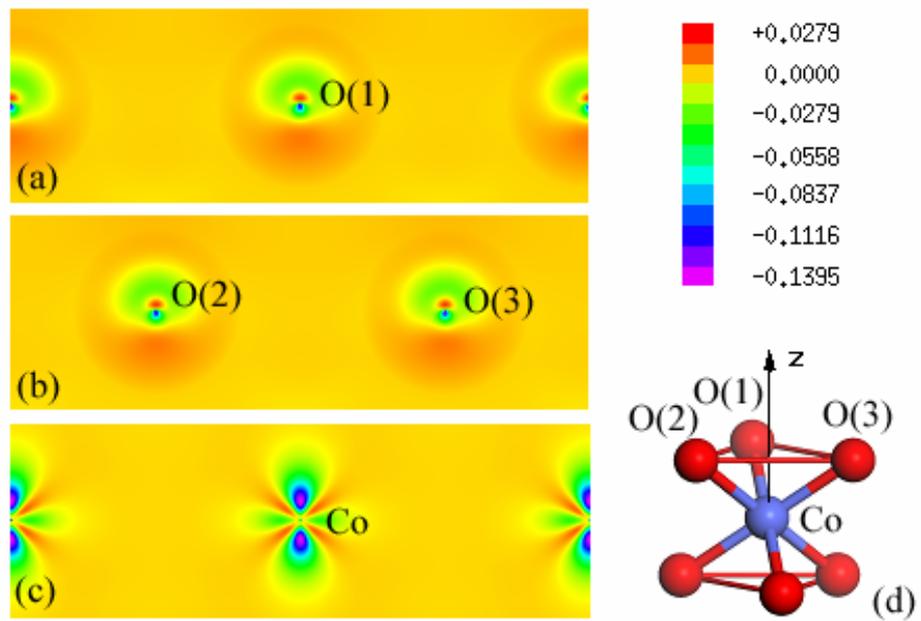

Fig. 1

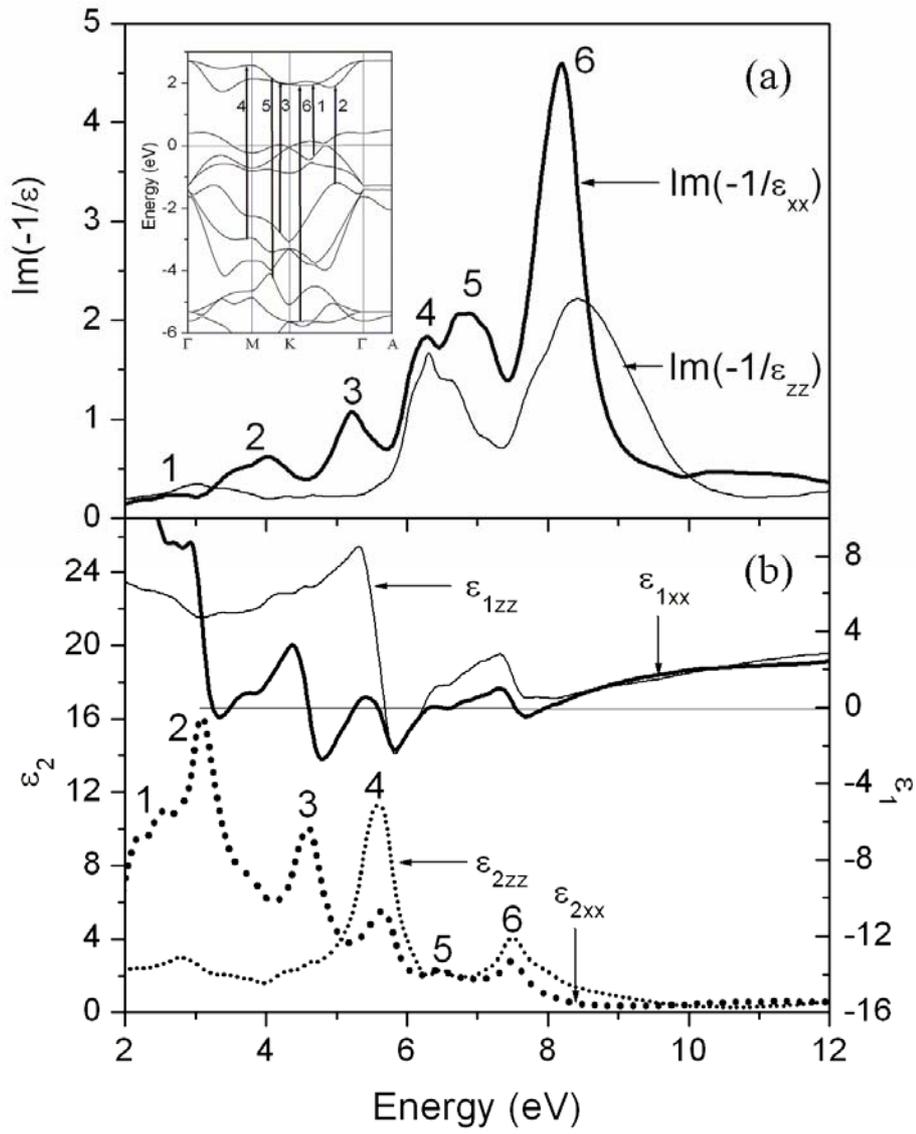

Fig. 2

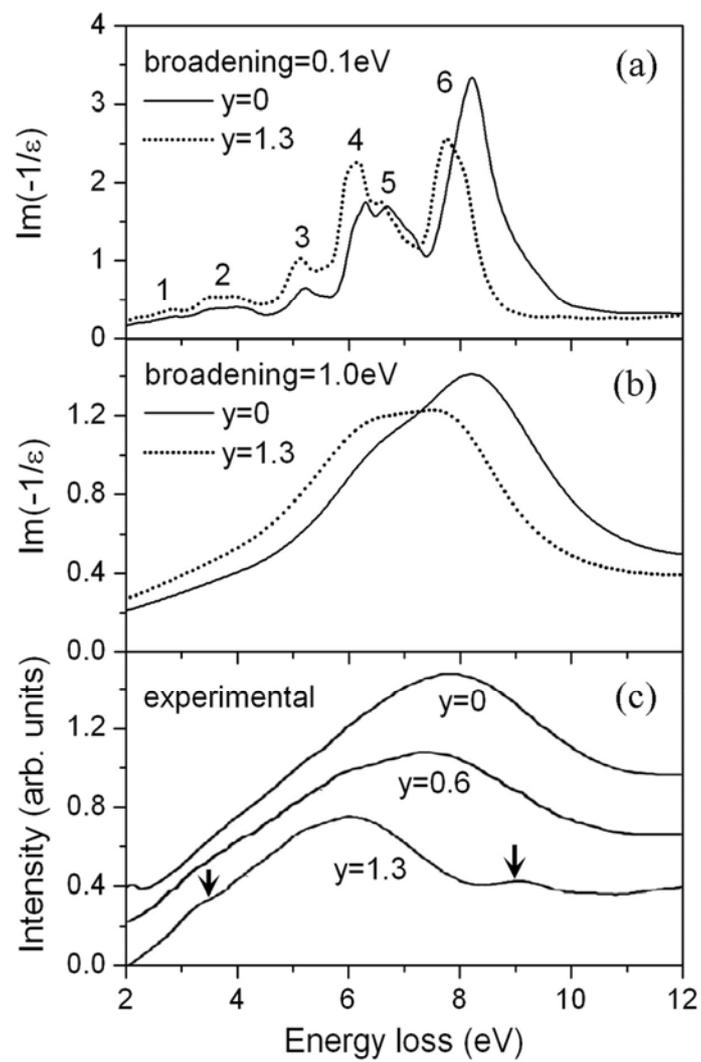

Fig. 3

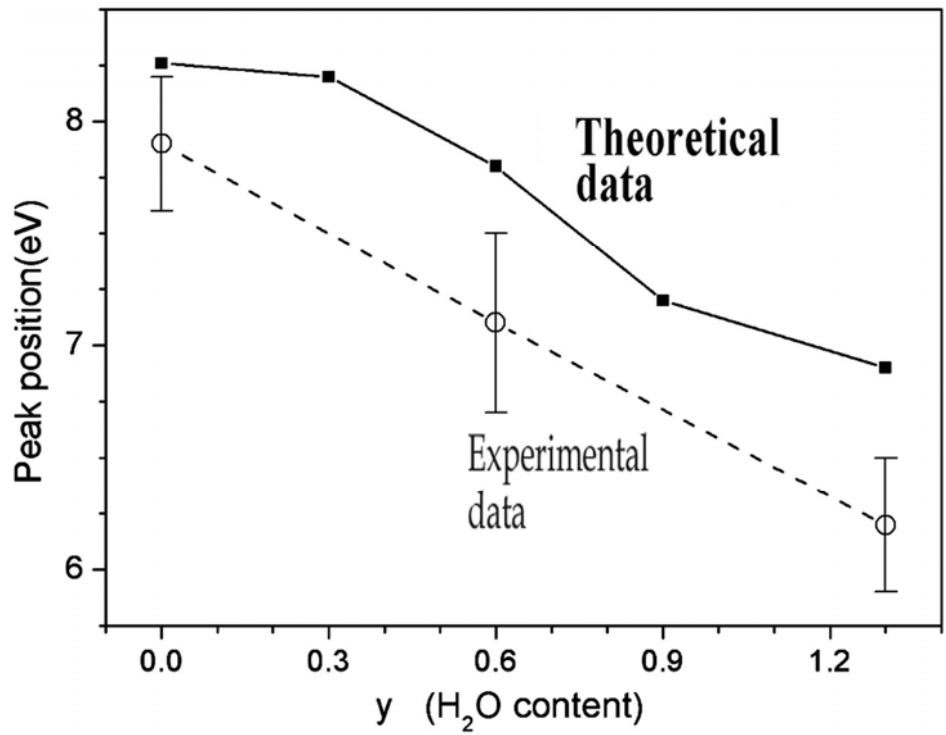

Fig. 4